\documentclass[pra,aps,twocolumn,showpacs,superscriptaddress]{revtex4}
\usepackage{epsfig}

\newcommand{\be}{\begin{equation}}
\newcommand{\ee}{\end{equation}}
\newcommand{\beq}{\begin{eqnarray}}
\newcommand{\eeq}{\end{eqnarray}}

\newcommand{\w}{\omega}
\newcommand{\W}{\Omega}
\newcommand{\g}{\gamma}

\begin{document}
\author{Vladimir A. Sautenkov}
\affiliation{
	Department of Physics and Institute for Quantum Studies,
	Texas A\&M University,
	College Station, Texas 77843-4242
}
\affiliation{
   Lebedev Institute of Physics, Moscow 119991, Russia
}

\author{Hebin Li}
\affiliation{
	Department of Physics and Institute for Quantum Studies,
	Texas A\&M University,
	College Station, Texas 77843-4242
}
\author{Yuri V.\ Rostovtsev}
\affiliation{
	Department of Physics and Institute for Quantum Studies,
	Texas A\&M University,
	College Station, Texas 77843-4242
}

%\author{Jonathan P. Dowling}
%\affiliation{
%Department of Physics and Astronomy, Louisiana State University, Baton Rouge, LA 70803-4001
%}

%\author{M.\ Suhail Zubairy}
%\affiliation{
%	Department of Physics and Institute of Quantum Studies,
%	Texas A\&M University,
%	College Station, Texas 77843-4242
%}

\author{Marlan\ O.\ Scully}
\affiliation{
	Department of Physics and Institute for Quantum Studies,
	Texas A\&M University,
	College Station, Texas 77843-4242
}
\affiliation{
Princeton Inst. for the Science and Technology of Materials
and Dept. of Mech. \& Aerospace Eng.,
Princeton University, 08544
}
\title{%An optical prism based on resonance ultra-dispersive media
Ultra-dispersive adaptive prism}

\begin{abstract}
We have %theoretically predicted and 
experimentally demonstrated 
an ultra-dispersive optical prism made from coherently driven Rb atomic
vapor. The prism possesses spectral angular dispersion that is six orders of
magnitude higher than that of a prism made of optical glass; 
it is the highest spectral angular dispersion that has ever been
shown (such angular dispersion allows one to spatially resolve light
beams with different frequencies separated by a few kHz). 
The prism operates near the resonant frequency of atomic vapor 
and its dispersion is optically controlled by a coherent driving field. 
\end{abstract}
%\pacs{42.50.Gy}
\date{\today}
\maketitle

\begin{figure}[tb] %fig2
\center{
\includegraphics[width=.7\columnwidth]{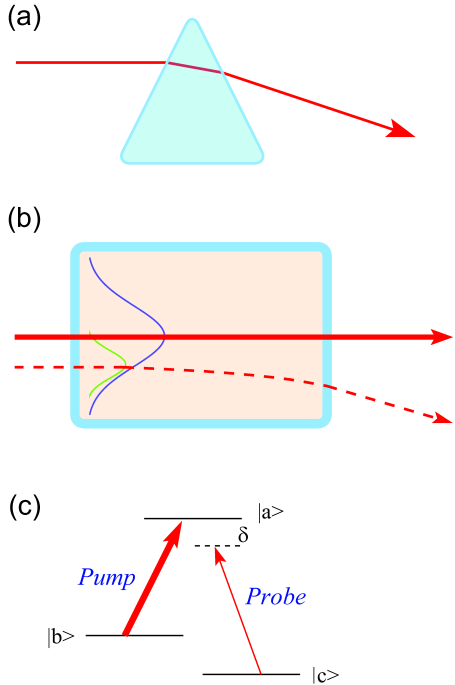} 
} 
\caption{\label{setup}
%Figure 1. 
(a) Refraction of light by the prism. 
(b) Configuration of the probe and control laser beams inside 
the cell of Rb vapor. 
One can see that our setup can be viewed as a super-high dispersive prism.
(c) Simplified scheme of the energy levels of Rb atoms. 
}
\end{figure}

A single frequency ray of light is bent by a prism upon an angle 
determined by the index of refraction, see Fig.1a. 
As was shown in \cite{newton}, the dispersion of 
the index of refraction leads to
spread of deviation angles for different light frequencies.

%Similar to the development of new meta-materials, 
%such as photonic crystals \cite{photon-crystal} or nanostructures \cite{nano},
Optical properties of matter, 
such as absorption, dispersion, and a variety of nonlinear characteristics, 
can be manipulated by electromagnetic fields 
\cite{thebook1,thebook2,thebook3}.  
For example, the applied coherent fields 
can eliminate absorption, enhance
the index of refraction \cite{mos-index,mos-index1,mos-index2},
induce chirality in nonchiral media~\cite{sau05prl}, 
produce usually forbidden forward Brillouin scattering 
or strong coherent backward scattering 
in ultra-dispersive resonant media~\cite{matsko,rost05prl}, 
slow down or speed up 
light pulses~\cite{slow-light,slow-light1,fast-light}, 
and the optical analog of Stern-Gerlach experiment~\cite{nature06}. 
Optically controlled giant nonlinearities may generate 
nonlinear signals using single photons~\cite{single-photon,harris05prl}. 
The enhanced nonlinearity can be employed for quantum information 
storage~\cite{lukin03science} and for manipulating of light propagating through
a resonant medium, such as 
stationary pulses of light in an atomic medium~\cite{lukin03nature}.

Here we experimentally demonstrate 
an ultra-dispersive prism (we refer to it as ``{\it a prism}''
because it deflects light, see Fig.1b) possessing the highest spectral
angular dispersion that has ever been experimentally observed (see Fig. 2,3).
The prism is made of a coherently driven atomic Rb 
vapor~\cite{thebook3} that has a spectral angular dispersion 
($d\theta/d\lambda = 10^3$ nm$^{-1}$) six orders of magnitude higher
than that of glass prisms ($d\theta/d\lambda = 10^{-4}$
nm$^{-1}$) or diffraction gratings ($d\theta/d\lambda =
10^{-3}$ nm$^{-1}$).

\begin{figure}[tb] %fig2
\center{
\includegraphics[width=.75\columnwidth]{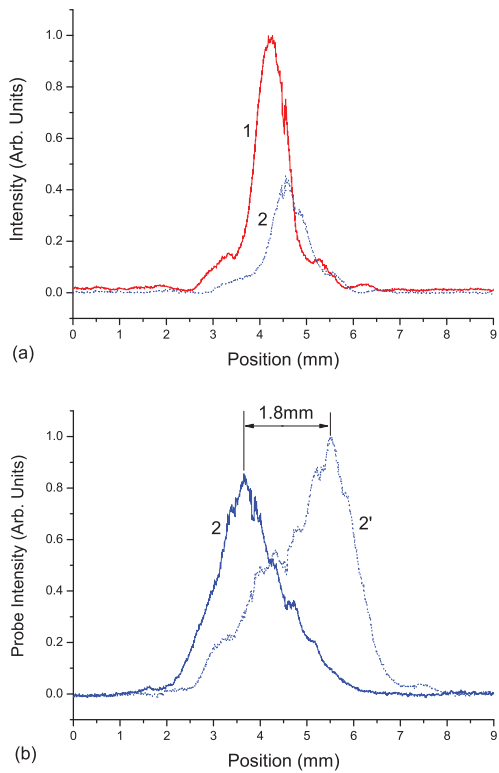}
} 
\caption{\label{1}
%Figure 1. 
(a) The spatial distributions of the control (1) and probe (2) 
fields at the input of the atomic cell. The probe is shifted to the right 
with respect to the control field. 
The spatial distributions of the probe fields (2) and (2') at the distance 
of 2.3 meters after passing the atomic cell
for different detunings corresponding to the maximal angles of deviation 
 (see Fig.~3a). 
}
\end{figure}

\begin{figure*}[tb] %fig2
\center{
\includegraphics[width=1.75\columnwidth]{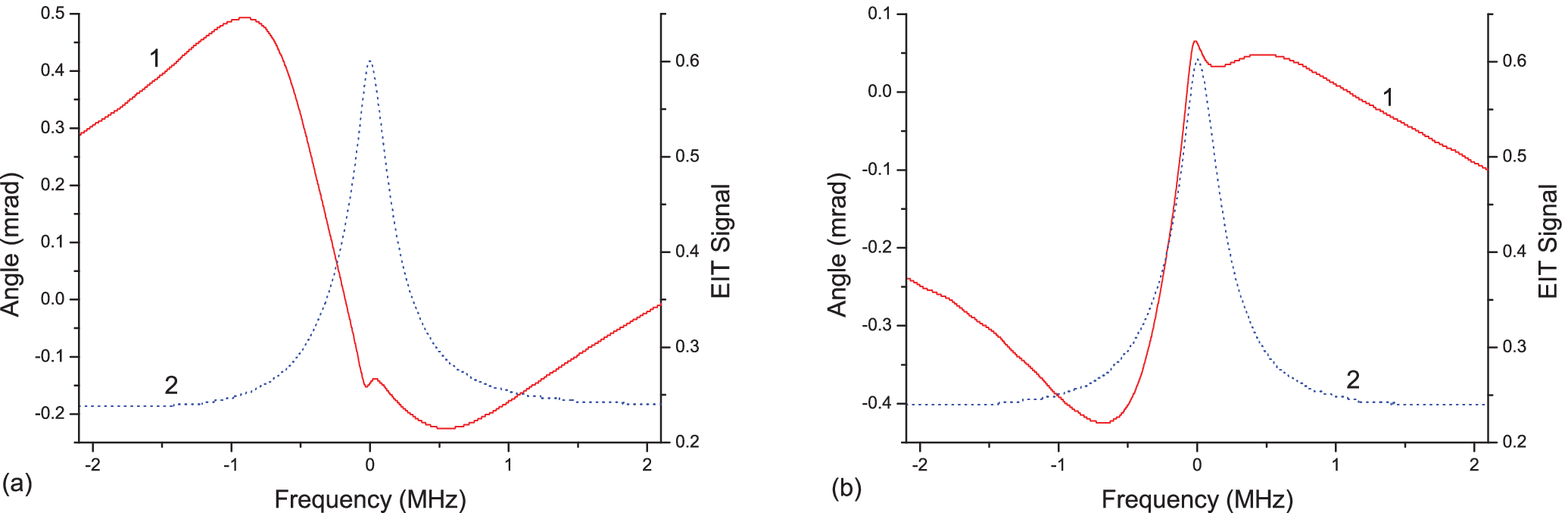}
} 
\caption{\label{2}
%Figure 1. 
(1) Dependence of the angle of the probe beam refraction on detuning for the 
probe beam initially shifted to right (a) and to the left (b) 
with respect to the control beam.  
(2) Dependence of the probe field transmission versus detuning. 
}
\end{figure*}

The physics of refraction of 
the ultra-dispersive coherently driven atomic medium is 
based on exciting quantum coherence. 
The wavevector $k$ depends on the light frequency as 
\be
k = {\w\over c} n,
\ee
where $n$ is the index of refraction. 
Assuming that the driving field has an inhomogeneous profile, 
then the index of refraction has a spatial gradient. 
The light ray trajectories  in an inhomogeneous medium 
can be found by solving an eikonal equation \cite{wolf} 
given by
\be
(\nabla \Psi)^2 = k^2 ={\w^2\over c^2} n^2
\ee
where $\Psi$ is the phase of electromagnetic wave.
%Residual absorption is the following
%$\kappa L = {\g_{cb}+\ds{\W^2/\G V_g}L \sim 1$, 
Then the light turning angle can be estimated as
\be
\theta \simeq L \nabla n. 
\label{gradient}
\ee
where $n = \sqrt{1 + 4\pi\chi(\w)}$, 
$L$ is the length of a medium and $\nabla n(\w)$ is the 
gradient of the index of
refraction in the direction perpendicular to propagation. 
The atomic susceptibility of coherently driven
medium $\chi(\w)$ \cite{thebook3} is given by
\be
\Re[\chi(\w)] = \eta\g_r\delta\w{\W^2 - \g_{cb}^2 - \delta\w^2\over
(\W^2 + \g_{cb}\g - \delta\w^2)^2 + \delta\w^2(\g_{cb} + \g)^2},
\ee
where $\eta= 3\lambda^3 N/16\pi^2$,  
$N$ is the density of Rb vapor, 
$\g$ is the relaxation rate at optical transition, $\g_{cb}$ is the
relaxation rate at the long-lived lower frequency (spin) transition, 
$\W$ is the Rabi frequency of control field, $\delta\w=\w - \w_{ab}$ is the
detuning of the probe field from atomic transition 
$\w_{ab} = 2\pi c/\lambda$; and 
$\lambda$ is the wavelength of resonant transition.
Then, for realistic parameters, such as $\delta\w \simeq 1$ kHz, 
$\g_{cb} = 1$ kHz, $N \simeq 10^{13}$ cm$^{-3}$, $L = 10$ cm
the estimate yields $\theta \simeq 0.1$, 
which shows a lot of potential for implementation of the predicted effect.
Note here that the spatial dependence of gradient of the driving field 
is important, and also that  
the effect can be increased even more by using an enhanced index of refraction
without absorption \cite{mos-index,mos-index1,mos-index2}. 

%\section{Experimental setup}

%Experimental setup 
The configuration of the laser beams is shown in Fig.1b, 
the practical details can be found in \cite{sau05pra}. 
The laser frequency is tuned to the center of the 
Doppler broadened D$_1$ line of Rb$^{87}$ (transition $F=2 - F=1$). 
Two orthogonally polarized beams, control ($P_c =0.5$ mW) and probe ($P_p
=0.5$ mW), create coherence between ground state Zeeman sublevels 
%Schematic $\Lambda$ scheme is 
as shown in Fig.~1c. A heated Rubidium cell 
($l =7.5$ cm, $N = 3\;\; 10^{11}$ cm$^{-3}$) 
is installed in a magnetic shield and 
two-photon detuning is varied by changing the magnitude of a 
longitudinal magnetic field. 
%After passed the cell ($l =7.5$ cm), the control and probe beams are
%separated. 

We employ two independent techniques to measure the probe beam position and 
the angle of deviation. 
The first technique is based on using a CCD camera and a 
removable mirror in front of the cell 
to measure the positions of the control and probe beams. 
The CCD camera is used to record an optical field distribution 
for selected two-photon detuning.  
In the second method, we use a position sensitive detector (PSD)
\cite{weis92ol} to accurately measure the beam direction versus
two-photon detuning. 
The distance from the center of the cell to PSD is 1 meter 
and to the CCD camera is 2.3 meters. 
Measurements by both techniques are consistent with each other.

Before the cell, the control and probe beams are parallel to each other. 
The probe beam can be adjusted to the left or to the right side of
the control beam profile by tilting a parallel glass plate. 
Then, after the cell, the probe and control beams are not longer parallel
(see Fig. 1b). When the probe beam is
shifted to the right side, as shown in Fig.~2a,
the observed probe beam profiles  
for two detunings are shown in Fig.~2b. 
The corresponding dependence of the angles of deviation on detuning is 
shown in Fig.~3a. The dependence corresponding to the shift of the probe field
to the left side of the control field is shown in Fig.~3b. One can see that
a different sign of the control field gradient changes the dependence on
the detuning. 

The width of the probe beam ($0.7$~mm at the Rb cell)
is increased twice at 2.3 meter distance from the cell due to diffraction
(the diffraction opening for a 
Gaussian beam profile is given by $2\lambda/\pi d$, where $d$ is the diameter
of the laser beam). For the data shown in Fig. 2b, the displacement
due to the ultra prism effect is larger than the spread of the probe beam due
to diffraction. 

In conclusion, we have experimentally demonstrated a EIT prism 
yielding large angular dispersion.  
The obtained results show the dependence of the angle of
deviation on the detuning that is introduced by a magnetic field. 
It follows from Eq.(\ref{gradient}), that the
angle of deviation is related to dispersion of the medium and the space 
gradient. Alternating the sign of the spacial gradient 
by shifting the probe beam, 
we can see the change of the dependence of the angle of deviation on the
two-photon detuning. 

The scheme holds promise for many applications.  
Such ultra-high frequency dispersion could be used for a compact
high spectral resolution spectrometer, similar
to compact atomic clocks and magnetometers \cite{hollberg}. 
The prism has a huge angular dispersion 
($d\theta/d\lambda = 10^3$ nm$^{-1}$)  
%that is the orders of magnetude higher
%than the typical prisms ($d\theta/d\lambda = 10^{-4}$ nm$^{-1}$, $R=10^4$), 
%diffraction gratings ($d\theta/d\lambda = 10^{-3}$ nm$^{-1}$, $R=10^6$), 
%or even interferometers ($R=10^9$) 
which can spatially resolve spectral widths of a few kHz
(%corresponding 
spectral resolution 
$R=\lambda/\delta\lambda\simeq 10^{12}$). 
We have observed the angle of deviation to be an order of
magnitude larger than the one previously observed in 
an inhomogeneous magnetic field \cite{nature06}.
We emphasis that the angle can be increased 
even further by using the enhanced index of refraction
without absorption \cite{mos-index,mos-index1,mos-index2}. 

The ability to control the direction of light
propagation by another light beam in transparent medium 
can be applied to optical imaging and to all-optical light steering
\cite{qingqing06proc}.
Also, this prism can be used for all-optical controlled delay lines
for radar systems. This technique can be easily extended to short pulses
by using the approach developed in \cite{qingqing05pra}. 

On the other hand, together with application to relatively intense classical 
fields, the ultra-dispersive prism can have
application to weak fields, such as a single photon source, and controlling 
the flow of photons at the level of 
a single quanta~\cite{single-photon,harris05prl}, 

We thank Hui Chen and M. Suhail Zubairy for useful discussions and 
gratefully acknowledge the support from 
the Defense Advanced Research Projects, the Office of Naval Research 
under Award No. N00014-03-1-0385, 
the Robert A.\ Welch Foundation (Grant \#A1261).


\begin{thebibliography}{99}

\frenchspacing

\bibitem{newton} Isaac Newton, {\it Opticks: Or a Treatise of the Reflections,
  Refrections, Inflections and Colours of Light}, (4th ed. London) 1730; 
Reprinted by (New York, Dover) 1952. 

%\bibitem{thebook} 
%O. A. Kocharovskaya and Ya. I. Khanin, 
%Coherent popultion trapping and the attendant effect of absorptionless
%propogation of ultrashort pulses trains in a 3-level medium,
%Pis'ma Zh. Eksp. Teor. Fiz. 48, 581 (1988)
%[JETP Lett. 48, 630 (1988)].

\bibitem{thebook1}
E. Arimondo, in Progress in Optics, 
edited by E. Wolf (Elsevier, Amsterdam, 1996), Vol. XXXV, pp. 257-354.
\bibitem{thebook2}
S. E. Harris, 
Electromagnetically induced transparency, 
Phys. Today 50, No. 7, 36 (1997).
\bibitem{thebook3}
M. O. Scully and M. S. Zubairy, {\it Quantum Optics},
  (Cambridge University Press, Cambridge, England, 1997).

%\bibitem{photon-crystal} %J.-M. Lourtioz, {\it Photonic crystals : towards
%  nanoscale photonic devices} (Berlin, Springer, 2005);
% Steven G. Johnson, John D. Joannopoulos, {\it Photonic crystals : the road
% from theory to practice} (Boston, Kluwer Academic Publishers, 2002).
% John D. Joannopoulos, Robert
%  D. Meade, Joshua N. Winn, {\it Photonic crystals: molding the flow of
%  light} (Princeton, N.J., Princeton University Press, 1995).

%\bibitem{nano} {\it Nanostructured materials and nanotechnology}, edited by
%  Hari Singh Nalwa, (San Diego, Calif., London, Academic Press, 2002).



\bibitem{mos-index}
M. O. Scully, 
Enhancement of the index of refraction via quantum coherence,
Phys. Rev. Lett. 67, 1855 (1991).

\bibitem{mos-index1} 
A. S. Zibrov, M. D. Lukin, L. Hollberg, D. E. Nikonov, M. O. Scully,
H. G. Robinson, and V. L. Velichansky, 
 Experimental Demonstration of Enhanced Index of Refraction via Quantum
 Coherence in Rb, 
Phys. Rev. Lett. 76, 3935 (1996).
\bibitem{mos-index2}
U. Rathe, M. Fleischhauer, S.-Y. Zhu, T. W. Hansch, and M. O. Scully,
 Nonlinear theory of index enhancement via quantum coherence and interference,
Phys. Rev. A 47, 4994 (1993).

\bibitem{sau05prl} V. A. Sautenkov, Y. V. Rostovtsev, H. Chen, P. Hsu,
  G. S. Agarwal, and M. O. Scully, 
Electromagnetically induced magnetochiral anisotropy in a resonant medium,
Phys. Rev. Lett. 94, 233601 (2005).

\bibitem{matsko} A. B. Matsko, Y. V. Rostovtsev, M. Fleischhauer, and
  M. O. Scully, 
Anomalous stimulated Brillouin scattering via ultraslow light,
Phys. Rev. Lett. 86, 2006-2009 (2001).

\bibitem{rost05prl} Y. Rostovtsev, Z.E. Sariyanni, M.O. Scully, 
Electromagnetically induced coherent backscattering,
Phys. Rev. Lett., accepted to be published (2006).


\bibitem{slow-light} A.B. Matsko, O. Kocharovskaya, Y. Rostovtsev, G.R. Welch,
A.S. Zibrov, M.O. Scully, 
Slow, ultraslow, stored, and frozen light, 
The advances in Atomic, Molecular, and Optical
Physics 46, 191 (2001), edited by B. Bederson and H. Walther.

\bibitem{slow-light1} R.W. Boyd, "Slow" and "fast" light
Prog. in Optics 43, 497 (2002).

\bibitem{fast-light} G.M. Gehring, 
A.  Schweinsberg, C. Barsi, N. Kostinski, R.W. Boyd,
Observation of backward pulse propagation through a
medium with a negative group velocity, Science 312, 895 (2006).

\bibitem{nature06} L. Karpa, M. Weitz, 
A Stern-Gerlach experiment for slow light,
Nature Physics 2, 332 (2006).

\bibitem{single-photon}
S. E. Harris and Y. Yamamoto,
Photon Switching by Quantum Interference,
Phys. Rev. Lett. 81, 3611 (1998).

\bibitem{harris05prl}
 V. Balic, D. A. Braje, P. Kolchin, G. Y. Yin, and S. E. Harris, 
Generation of Paired Photons with Controllable Waveforms,
 Phys. Rev. Lett. 94, 183601 (2005)

\bibitem{lukin03science}
C.H. van der Wal, M.D. Eisaman, A. Andre, R.L. Walsworth,
D.F. Phillips, A.S. Zibrov, M.D. Lukin, 
Atomic memory for correlated photon states 
Science 301, 196 (2003).

\bibitem{lukin03nature}
M.Bajcsy, A.S.Zibrov and M.D.Lukin, 
Stationary pulses of light in an atomic medium,
Nature (London) 426, 638 (2003).


\bibitem{wolf}  Max Born and Emil Wolf, 
{\it Principles of optics : electromagnetic theory of propagation, interference and diffraction of light}, 
(Cambridge, UK ; New York ; Cambridge University Press, 1997).

\bibitem{sau05pra} V. A. Sautenkov, Yu. V. Rostovtsev, and M. O. Scully,
Switching between photon-photon correlations and Raman anticorrelations in a
coherently prepared Rb vapor, 
Phys. Rev. A 72, 065801 (2005).

\bibitem{weis92ol} R. Schlesser, A. Weis, 
Lifgt-beam deflection by cesium vapor in a transferse-magnetic field, 
Opt. Lett. 17, 1015 (1992).

\bibitem{qingqing05pra} Q. Sun, Y. Rostovtsev, J. Dowling, M.O. Scully,
M.S. Zubairy, 
Optically controlled delays for broadband pulses,
Phys. Rev. A72, 031802 (2005).

\bibitem{qingqing06proc} Q. Sun, Y. V. Rostovtsev, and M. S.
  Zubairy, 
All optically controlled steering of light,
Proc. SPIE 6130, 61300S (2006).


\bibitem{hollberg} S. Knappe, P.D.D. Schwindt, V. Gerginov, V. Shah, 
L. Liew, J. Moreland, H. G. Robinson, L. Hollberg, J. Kitching,
Microfabricated atomic clocks and magnetometers, 
J. Opt. A: Pure Appl. Opt. 8, S318 (2006).

\end{thebibliography}
\end{document}